\documentclass{aa}  
\usepackage{graphicx}
\usepackage{txfonts}
\newcommand{\xmm}{{XMM-{\em Newton} }}
\newcommand{\detml}{{DET\_ML }}
\begin{document}
   \title{The first XMM-Newton slew survey catalogue: XMMSL1}

   \subtitle{}

   \author{R.D. Saxton
          \inst{1}
          \and
          A.M. Read\inst{2}
          \and
          P. Esquej\inst{3,1}
          \and
          M.J. Freyberg\inst{3}
          \and
          B. Altieri\inst{1}
          \and
          D. Bermejo\inst{1,4}
          }

   \offprints{R. Saxton}

   \institute{XMM SOC, ESAC, Apartado 78, 28691 Villanueva de la Ca\~{n}ada, Madrid
              , Spain\\
              \email{richard.saxton@sciops.esa.int}
         \and
             Dept. of Physics and Astronomy, University of Leicester, Leicester LE1 7RH, U.K.
         \and
             Max-Planck-Institut f\"{u}r extraterrestrische Physik, D-85748 Garching, Germany
         \and
             Instituto de Astrofísica de Andalucía, CSIC, Granada, Spain
        }

   \date{Received September 15, 1996; accepted March 16, 1997}

 
  \abstract
   {}
   {We report on the production of a large area, shallow, sky survey, from \xmm slews.
The great collecting area of the mirrors coupled with the high
quantum efficiency of the EPIC detectors have made \xmm
the most sensitive X-ray observatory flown to date. We use data taken
with the EPIC-pn camera during slewing manoeuvres to perform an 
X-ray survey of the sky.  }
   { Data from 218 slews have been subdivided into small images and source searched.
   This has been done in three distinct energy bands; a soft (0.2-2 keV)
   band, a hard (2-12 keV) band and a total \xmm band (0.2-12 keV). Detected
   sources, have been quality controlled to remove artifacts and a
   catalogue has been drawn from the remaining sources.}
   {A 'full' catalogue, containing 4710 detections and a 'clean' catalogue containing 2692 sources have been produced, from 14\% of the sky. In the hard X-ray band (2-12 keV) 257 sources are detected in the clean catalogue to a flux limit of $4\times10^{-12}$ ergs s$^{-1}$ cm$^{-2}$. The flux limit for the soft
(0.2-2 keV) band is $6\times10^{-13}$ ergs s$^{-1}$ cm$^{-2}$ and for the
total (0.2-12 keV) band is $1.2\times10^{-12}$ ergs s$^{-1}$ cm$^{-2}$. The source positions are shown to have an uncertainty of 8\arcsec (1~$\sigma$ confidence).} 
   {}

   \keywords{X-rays: general -- surveys --
               }

   \maketitle
%

\section{Introduction}

There is a strong tradition in X-ray astronomy of using data taken during
slewing manoeuvres to perform shallow surveys of the sky. 
The Einstein (\cite{elvis}), Exosat (\cite{reynolds}) and RXTE (\cite{revnivtsev}) slew surveys all provide a useful 
complement to dedicated all-sky surveys such as ROSAT (\cite{voges}) and
HEAO-1 (\cite{piccinotti})  and the
smaller area, medium sensitivity ASCA survey (\cite{ueda}) and pencil-beam \xmm 
(\cite{hasinger}) and Chandra (\cite{brandt}) deep looks. It was long recognised that 
XMM-Newton (\cite{jansen}) with its great collecting area, efficient CCDs,
wide energy band and tight point-spread function (PSF) has the potential to 
make an important contribution to our knowledge of the local universe
from its slew data. Early estimates (\cite{Lumb98}; \cite{jonesandlumb}), based on an expected slewing speed of 30 degrees per hour, 
and a slightly lower
background level than that actually encountered in orbit, predicted a 0.5--2 keV
flux limit of $2\times10^{-13}$ erg cm$^{-2}$s$^{-1}$. While the chosen in-orbit
slew speed of 90 degrees per hour reduces the sensitivity,
 initial assessments of the data showed that the quality of the data is 
good and that many sources are detected (\cite{freyberg}). A review of properties shows that the
\xmm slew survey compares favourably with other large area surveys in terms of depth and positional accuracy (Table 1).

During slews all the three imaging EPIC cameras take data in the observing mode
set in the previous pointed observation and with the Medium filter in
place. The slew speed of 90 degrees per hour combined with the slow
readout time of the MOS detectors (2.6s; \cite{turner}) means that 
sources appear as 
long streaks in the MOS cameras but are 
well formed in the fast observing modes of the pn camera (\cite{struder}). 
For this reason, only the EPIC-pn data have been analysed.

In this paper we present a catalogue drawn from slews taken between revolutions
314 and 978 covering a sky region of 6240 square degrees. The main 
properties of the slew survey discussed in this paper are given in Table 2. 

\begin{table}
\caption{Properties of large area X-ray surveys}
\label{table:SurveySumm}      
\begin{center}
\begin{tabular}{l c c c c}        
\hline\hline                 
Satellite & Energy range & Coverage$^{a}$ & Flux lim. & Position \\    
          & (keV) &    \% of sky   &  $^{b}$      &  error \\    
\hline                        
   RASS & 0.2-2.4 & 92 & 0.03 & 12\arcsec \\      
   Einstein slew & 0.2-3.5 & 50 & 0.3    & 1.2\arcmin \\
   {\bf XMM slew (soft)} & {\bf 0.2-2} & {\bf 14} & {\bf 0.06}  & {\bf 8\arcsec}  \\
\hline                        
   EXOSAT slew & 1-8 & 98 & 3  & 20\arcmin \\
   HEAO-1/A2 & 2-10 & 100 & 3 & 60\arcmin \\      
   RXTE slew & 3-20 & 95 & 1.8  & 60\arcmin \\
   {\bf XMM slew (hard)} & {\bf 2-12} & {\bf 14} & {\bf 0.4}  & {\bf 8\arcsec} \\
\hline                                   
\end{tabular}
\\
\end{center}
$^{a}$ The \xmm slew sky coverage has been computed by adding the area
contained in all of the images used in source searching with an exposure time
greater than 1 second. \\ 
$^{b}$ Flux limit, units of $10^{-11}$ ergs $s^{-1}$ cm$^{-2}$ \\
\end{table}

\begin{table}
\caption{Properties of the \xmm slew survey}
\label{table:slewprops}      
\begin{center}
\begin{tabular}{l c c c}        
\hline\hline                 
Property & \multicolumn{3}{c}{Range (keV)}\\    
         & 0.2--2 & 2--12 & 0.2--12  \\    
\hline                        
Observing time (s) & $5.4\times10^{5}$ & $5.4\times10^{5}$ & $5.4\times10^{5}$ \\
Mean exposure time$^{a}$ & 6.2 s & 6.0 s & 6.1 s\\
Total number of photons & $1.6\times10^{6}$ & $2.2\times10^{6}$ & $3.8\times10^{6}$ \\
Mean background$^{b}$ & 0.09 & 0.14 & 0.23\\
Median source count rate$^{c}$ & 0.68 & 0.90 & 0.81\\
Limiting source flux$^{d}$ & $6.0\times10^{-13}$ & $4.0\times10^{-12}$ & $1.2\times10^{-12}$\\
Num. sources (full cat)$^{e}$ & 2606 & 692 & 3863\\
Num. sources (clean cat)$^{e}$ & 1874 & 257 & 2364\\

\hline                                   
\end{tabular}
\\
\end{center}
$^{a}$ The mean exposure time, after correcting for the energy-dependent
vignetting.\\
$^{b}$ cnts/arcmin$^{2}$. \\
$^{c}$ cnts/second. \\
$^{d}$ ergs/s/cm$^{2}$. Based on a detection of 4 photons from a source passing near the detector centre (exp. time = 10s), with a power-law spectrum of slope 1.7 and galactic absorption of 3$\times10^{20}$ atoms cm$^{-2}$. \\
$^{e}$ The number of sources flagged as good in the full and clean catalogues (see text). \\
\end{table}

Slews have been source searched down to a likelihood threshold of 8, which 
after manual rejection of false detections gives 4710 candidate sources. Using simulations we have been able to identify a subset of high significance sources, with a likelihood
threshold dependent upon the background conditions, that gives 2692 sources
with a spurious fraction of $\sim4\%$ (the "clean" catalogue). Of these, 
2621 are from unique sources. In the hard (2--12 keV) band the clean catalogue
contains 257 sources (253 unique) of which $\sim9\%$ are expected to be due to 
statistical fluctuations.

The slew catalogue and accompanying images and exposure maps have been 
made available through the XMM Science Archive (XSA) as a queryable
database and as FITS files\footnote{The catalogue was initially released
on May 3 2006. In this paper we discuss the updated version released in October
2006.}. A summary of scientific highlights from the slew survey has been 
published in \cite{Read06}.


%
\section{Data selection}
The \xmm satellite moves between targets by performing
an open-loop slew along the roll and pitch axes
and a closed-loop slew, where  measurements from the star tracker are used in addition to the
Sun--sensor measurements to provide a controlled slew about all three axes, to correct for residual errors in the
long open-loop phase. The open-loop slew is performed at a steady rate
of about 90 degrees per hour and it is data from this phase which may be
used to give a uniform survey of the X-ray sky.
                                                                            
\begin{figure}
\begin{center}
\begin{tabular}{c}
\rotatebox{-90}{\includegraphics[height=9cm]{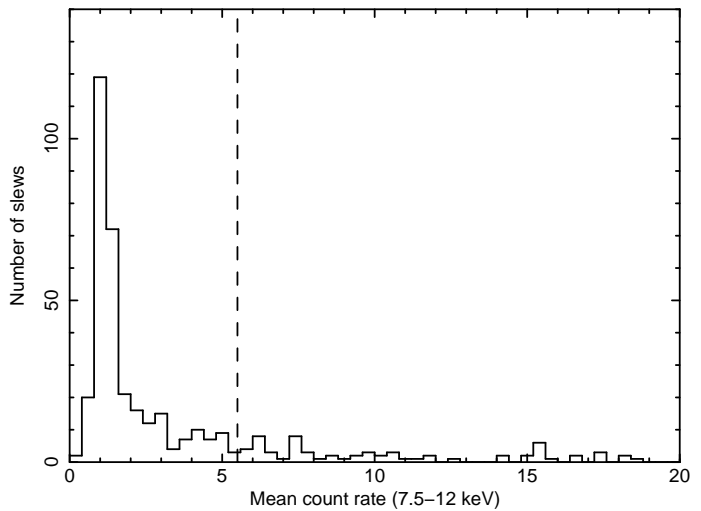}}
\end{tabular}
\end{center}
\caption[Slew background histogram]
{ \label{fig:bckhist} The background level distribution measured from the
mean 7.5--12 keV count rate in each slew. The dashed line shows the
cut-off used to exclude high background slews in the tail of the distribution. }
\end{figure}

\begin{figure}
\begin{center}
\begin{tabular}{c}
\rotatebox{-90}{\includegraphics[height=9cm]{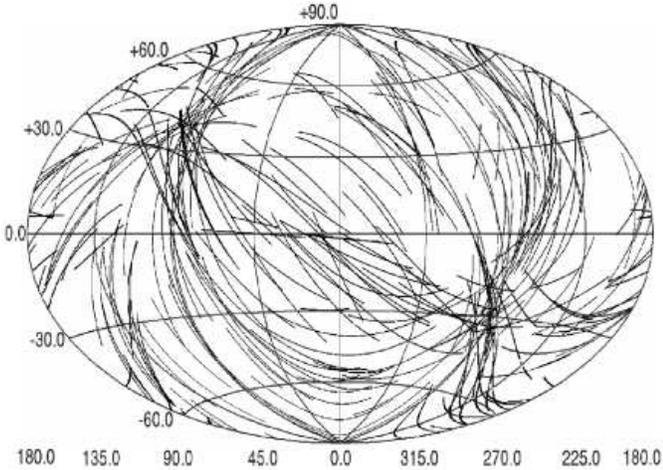}}
\end{tabular}
\end{center}
\caption[Slew paths in celestial coordinates]
{ \label{fig:slewpath} The paths of the slews used in the construction of
XMMSL1 in Galactic coordinates.}
\end{figure}

Slew Data Files (SDF) have been stored in the XSA
from revolution 314 onwards (before this date slews were performed
with the CLOSED filter in place and no scientifically useful data
was taken). Data from revolutions 314 to 978 have been
used for this first catalogue but slews continue to be accumulated
and increments to the catalogue are planned to be released on a regular
basis.
Only science data from slews with a duration of greater than 30 minutes 
were downlinked during the majority of these revolutions\footnote{This policy was changed from revolution 921 onwards to include all slews longer than 15 minutes}.
For the {\em Slew Survey} catalogue
we have selected only EPIC-pn exposures
performed in {\em Full Frame} (FF),
{\em Extended Full Frame} (eFF), and {\em Large Window}
(LW) modes, i.e.\ modes where all 12 CCDs are
integrating (in LW mode only half of each CCD).
The corresponding cycle times are
73.36\,ms, 199.19\,ms, and 47.66\,ms, which converts
to a scanned distance of 6.6 arcseconds, 17.9 arcseconds, and
4.3 arcseconds per cycle time, respectively for a slew speed of 90 degrees
per hour.
In the {\em Small Window} mode only the
central CCD is operated and a window of $64\times64$
pixels is read out, i.e.\ only about 1/3 of the single, prime CCD.
In the fast modes, {\em Timing} and {\em Burst},
only 1-dimensional spatial information
for the central CCD is available
and thus these modes are not well suited for
source detection. It was discovered in initial tests that slews
with a high background gave a large number
of false detections and resulted in extremely long execution times
for the source searching software. For this reason, slews with 
 an average 7.5--12 keV (FLAG=0, PATTERN=0-4) count rate in 
excess of 5.5 c/s (25\% of all slews), were
excluded from the analysis (Fig.~\ref{fig:bckhist}). This leaves
a nominal 312 slews potentially useful for scientific analysis.
In practice a significant number of slews were not able to be processed
for a variety of reasons including failure to create exposure maps, 
unreasonably high exposure times, attitude reconstruction problems, 
missing keywords and the processing producing too large images. 
While it is strongly hoped that some of these 
datasets may be recoverable in the future by improvements to the processing
system, for the purposes of this first catalogue they have been left out
and the catalogue constructed from 218 slews. A list of the observation numbers of these slews is available at 
{\it http://xmm.esac.esa.int/external/xmm\_science/slew\_survey}
{\it/obsid\_tab.html}
and the slew paths are shown in Fig.~\ref{fig:slewpath}.
 About 4\% of the covered area has been 
slewed over more than once and eventually a deeper survey 
will be available by combining overlapping slew data, especially near the ecliptic poles.

\begin{figure}
\begin{center}
\begin{tabular}{c}
\rotatebox{-90}{\includegraphics[height=8cm]{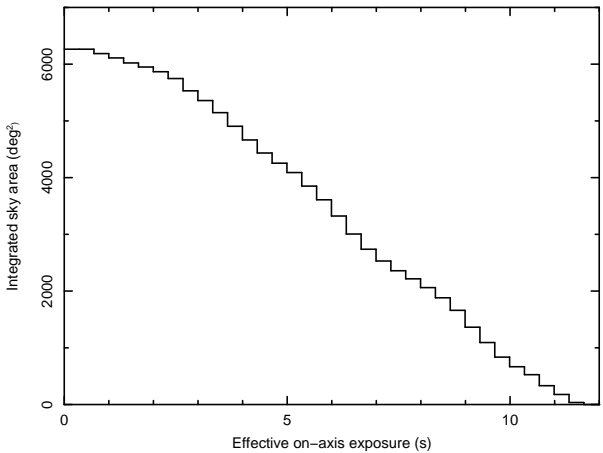}}
\end{tabular}
\end{center}
\caption[Exposure time v Sky area]
{ \label{fig:expsky} A histogram of the cumulative sky area covered
for a given effective on-axis exposure time in 
the total (0.2-12 keV) band.
}
\end{figure}

The EPIC-pn detector passes over a source in about 14 seconds depending
on the position of the source and the angle subtended by the Y-axis of the 
detector to the slew path (the impact angle). Normally this is close to zero, 
but an impact angle of up to 20 degrees is possible. If the source
passes through the detector optical-axis an effective on-axis exposure time of 
$\sim 11$ seconds is achieved. In Fig.~\ref{fig:expsky} we show the exposure time of the
slew survey as a function of sky coverage. Due to the 
vignetting function, the mean exposure time is energy-dependent being
6.2 seconds in the soft (0.2--2 keV) band and 6.0 seconds in the hard (2--12 keV) band. 
Small ripples in the histogram are caused by gaps between 
the EPIC-pn CCDs
and also by the different observing modes; in LW mode only
half of the CCD is exposed and the maximum effective exposure time 
is ~6 seconds.
                                                                            
\section{Data processing}
The data have been used to perform three independent surveys; a soft band
(0.2--2 keV) X-ray survey with strong parallels to the ROSAT all--sky survey (\cite{voges}; RASS), a hard band (2--12 keV) survey and an \xmm full-band
(0.2--12 keV) survey.

Data reduction was performed as detailed below, with the
public \xmm Science Analysis Software (SAS), version 6.1  plus the following modifications:

\begin{itemize}
\item a modification for
the {\tt oal} library to handle the relatively large time gaps in the
Raw Attitude File (RAF); subsequently released with
{\tt SAS v7.0}.
\item an increase in the maximum number of attitude points which may be used
by {\tt eexpmap}; released with {\tt SAS v6.5}. 
\end{itemize}

\subsection{Initial reduction}
The \xmm Slew Data Files (SDFs) for EPIC-pn
were processed using the {\tt epchain} package of the SAS.
For diagnostic reasons a few parameters were set to
non-default values (e.g. events below 150\,eV were kept).

\subsection{Slew division}
Photon events are recorded initially in RAW or detector coordinates and have to
be transformed, using the satellite attitude history, into sky coordinates.
The tangential plane geometry commonly used to define a coordinate grid for
flat images is only valid for distances of 1--2 degrees from a reference
position, usually placed at the centre of the image. To avoid this
limitation, slew datasets have been divided into roughly one degree
by half a degree 
event files and attitude corrected using the task {\tt attcalc}. Images and exposure maps were then 
extracted from the event files using the tasks {\tt evselect} and {\tt eexpmap}.
This procedure relies on the attitude history of the satellite being accurately known
during the slew; a point which is addressed in section~\ref{sect:attitude} .
Software, based on SAS and ftools (http://heasarc.gsfc.nasa.gov/ftools ; \cite{blackburn}), to perform the procedure of 
dividing and attitude-correcting slew data has been made available
via the \xmm web-site\footnote{http://xmm.esac.esa.int/sas}. 

The procedure has been repeated in three 
separate energy bands: full band (0.2$-$0.5\,keV [pattern=0] +
0.5$-$12.0\,keV [pattern=0$-$4]), soft band (0.2$-$0.5\,keV
[pattern=0] + 0.5$-$2.0\,keV [pattern=0$-$4]), and hard band
(2.0$-$12.0\,keV [pattern=0$-$4]). 
During the data processing stage severe problems with the transfer and
storage of files from long slews were encountered. To alleviate this, 
only exposure maps for the full energy band were produced.
This leads to an approximation being needed for exposure times in the different
energy bands which is addressed in section~\ref{sect:countrates}.

\subsection{Source searching}
Pilot studies were performed to investigate  the optimum
processing and source-search strategies. 
Uneven (and heightened) slew exposure is
observed at the end of some slews (the 'closed-loop' phase) and images
with an exposure time greater than 20 seconds have been removed to 
ensure the uniformity of the survey. We
tested a number of source-searching techniques and found that the
optimum strategy was the usage of a semi-standard
`eboxdetect (local) + esplinemap + eboxdetect (map) + emldetect'
method, tuned to $\sim$zero background, and performed on a single
image containing just the single events (pattern=0) in the
0.2$-$0.5\,keV band, plus single and double events (pattern=0$-$4) in
the 0.5$-$12.0\,keV band. This is similar to the technique used for
producing the RASS catalogue (\cite{Cruddace88}) and resulted in the
largest numbers of
detected sources, whilst minimising the number of spurious sources
due to detector anomalies (usually caused by non-single, very soft
($<$0.5\,keV) events). The source density was found to be $\approx$0.5
sources per square degree to an emldetect detection likelihood threshold
(\detml) of 10 ($\sim3.9\sigma$).

\section{Attitude reconstruction and positional accuracy}
\label{sect:attitude}
                                                                            
The good point spread function of the X-ray telescopes
(\cite{aschenbach}) should allow slew source positions to be determined to an
accuracy of around 4 arcseconds, similar to that found for faint
objects in the 1XMM catalogue of serendipitous sources detected in pointed
observations \footnote{The First XMM-Newton Serendipitous Source Catalogue,  XMM-Newton Survey Science Centre (SSC), 2003.}. Any errors in the attitude reconstruction
for the slew could seriously degrade this performance and a major technical
challenge of the data processing has been to achieve the nominal accuracy. 

The attitude information of the XMM-Newton satellite is provided by
the Attitude and Orbit Control Subsystem (AOCS). A star tracker co-aligned
with the telescopes allows up to a maximum of five stars to be continuously
tracked giving accurate star position data every 0.5 seconds, which
operates in addition to the Sun
sensor that provides a precise Sun-line determination. Such information is
processed resulting in an absolute accuracy of the reconstructed astrometry
of typically 1 arcsecond (1 sigma) for pointed observations. 
For the open-loop slews, large slews outside the
star-tracker field of view of 3 x 4 degrees, the on-board software generates a
three axis momentum reference profile and a two-axis (roll and pitch)
Sun-sensor profile, both based on the ground slew telecommanding. During
slew manoeuvring a momentum correction is superimposed onto the reference
momentum profile and, as there are no absolute measurements for the yaw axis,
a residual yaw attitude error exists at the end of each slew that may be
corrected in the final closed-loop slew (Elfving 1999).
                                                                            
Two types of attitude data may be used as the primary
source of spacecraft positioning during event file processing.
They are the Raw Attitude File (RAF) and the Attitude History File (AHF).
For pointed observations, the RAF provides the attitude information at the maximum possible rate,
with one entry every 0.5 seconds while the AHF is a smoothed and filtered version
of the RAF, with times rounded to the nearest second. In slew datasets the
RAF stores attitude information every 40--60 seconds while the AHF
contains the same records as the RAF with identical positions and again
with timing information in integer seconds.
The user can select which one to use for data processing by
setting an environment variable.
                                                                            
In a pilot study where the AHF was used for attitude reconstruction,
source detection was performed and correlations with the ROSAT and 2MASS
catalogues indicated a slew relative pointing accuracy of $\sim10$ arcseconds
. However, an absolute
accuracy of 0-60 arcseconds (30 arcseconds mean) was obtained in the slew direction,
resulting in a thin, slew-oriented error ellipse around each source.
This error appears to be consistent with the error introduced by the
quantisation of the time to 1 second in the attitude file and led us to change the processing software
as a better accuracy should be obtained. As a test, the
RAF was used to compute the astrometry for some observations. Here
, an offset of $\sim1$ arcminute from the ROSAT positions was found,
but with a smaller scatter compared with the positions returned by the AHF
processing. The consistency of these offsets suggested that they could be due to a timing issue. After discussions with the flight dynamics group it 
was realised that the star tracker CCD integration time of  0.75 seconds
is not included in the times in the attitude history.
When this 0.75 seconds is subtracted from every entry in the
RAF we obtain an optimal attitude file for the processing.
Note that this offset remains in all XMM raw attitude files but an
automatic correction has been applied in the SAS software from version 7 
onwards. Also note that this discrepancy has no practical effect on normal 
stable-pointed observational data.                                 

\begin{figure}
\begin{center}
\begin{tabular}{c}
\rotatebox{90}{\includegraphics[height=9.5cm]{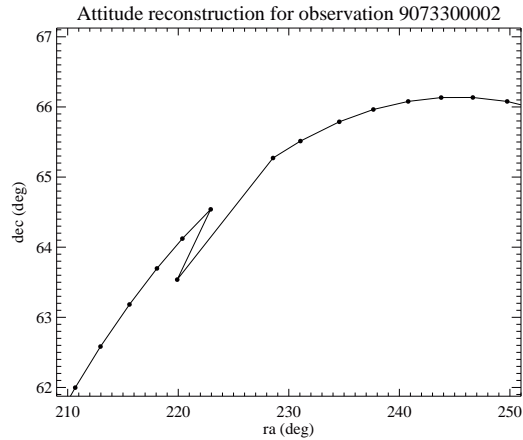}}
\end{tabular}
\end{center}
\caption[Rev 0733 peak]
{ \label{fig:peak}
A zoom into the problematic region of the attitude file in the slew 9073300002.
 The points show the generated attitude information.
}
\end{figure}

\begin{figure}
\begin{center}
\begin{tabular}{c}
\rotatebox{90}{\includegraphics[height=9.5cm]{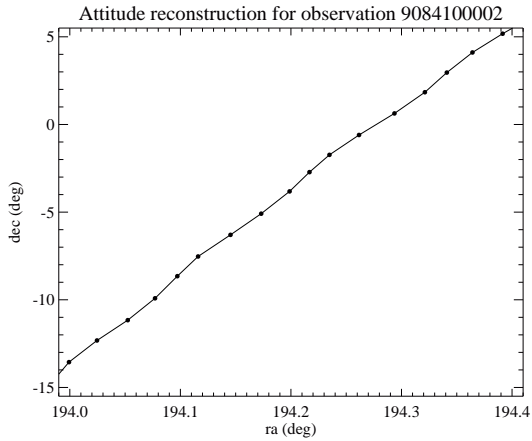}}
\end{tabular}
\end{center}
\caption[Rev 0841 turbulence]
{ \label{fig:turbulence}
A plot showing non-smooth, or turbulent, attitude reconstruction in
the revolution 0841 attitude file.}
\end{figure}

Other issues affecting the astrometry performance appeared after a careful
visual examination of the RAF files, where two types of peculiarities
appeared in some of the slews affecting either a localised region or the totality
of the slew. 
Five of the slews presented sharp discontinuities in the attitude
reconstruction, revealing the
existence of single bad RAF point.
As an example
a source in the slew 9073300002 was discovered to have a closest ROSAT counterpart at
a distance of 8 arcminutes. Investigation showed that the source was observed at a time
coincident  with a large error in the attitude file (Fig.~\ref{fig:peak}). 
A test involving the removal of the bad point and recalculation of the 
attitude, while showing an improvement in source positions didn't improve
the astrometry to the level of the other slews. Therefore, sources falling
in a region of bad attitude have been flagged in the catalogue
with the 'Position Suspect'
flag (see section~\ref{sect:flags}).
Sections of seven other slews displayed an attitude reconstruction 
that can best be described as turbulent 
 (Fig.~\ref{fig:turbulence}). Again, sources falling in these slew
sections have been marked with the 'Position Suspect'
flag. 

\begin{figure}
\centering
\rotatebox{90}{\includegraphics[height=9.5cm]{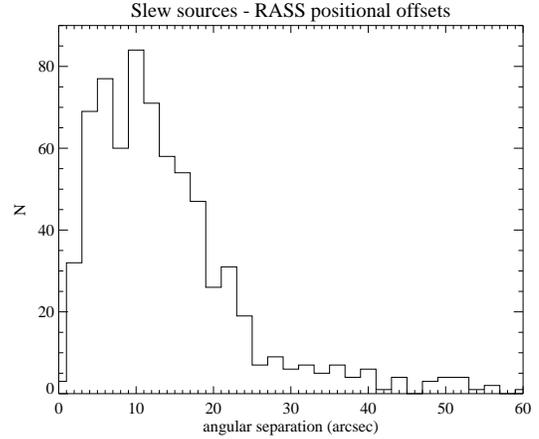}}
\rotatebox{90}{\includegraphics[height=9.5cm]{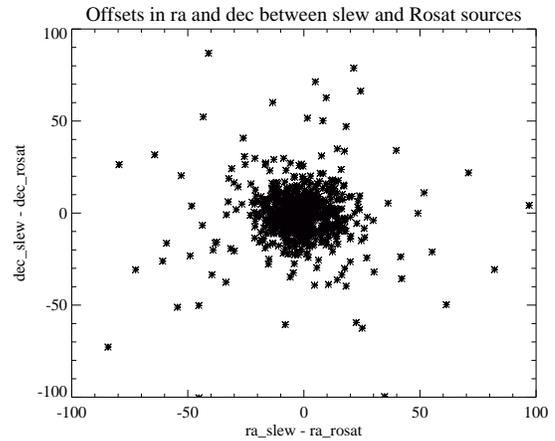}}
\caption[RASS comparison]
{ \label{fig:ROSAThist}
A comparison of slew source positions with those from
the RASS catalogue; 68\% of the sources lie within 16 arcseconds.
The upper panel shows a histogram of the offset magnitude while the
lower panel gives the absolute offset in arcseconds of the slew source
from the ROSAT position.}
\end{figure}

\begin{figure}
\begin{center}
\begin{tabular}{c}
\rotatebox{90}{\includegraphics[height=9.5cm]{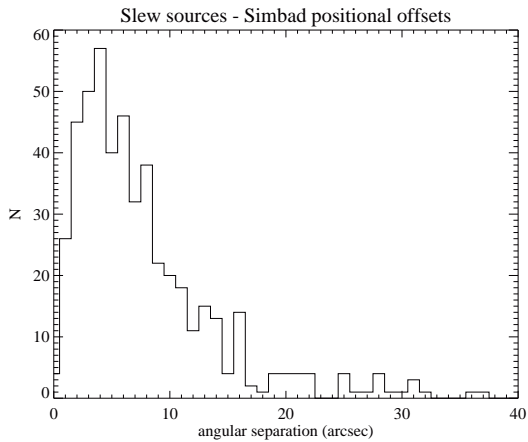}}
\end{tabular}
\end{center}
\caption[Simbad cross-correlation]
{ \label{fig:Simbadhist}
A histogram of the distribution of the angular separation 
    of the slew sources from their Simbad counterpart; 
    68\% of the matches lie within 8 arcseconds.}
\end{figure}

A subsample of 1260 non-extended sources (defined as having an extent parameter
$<2$ from the emldetect source fitting) with $\detml>10$, have been correlated with
several catalogues within a 60 arcseconds offset.
The correlation with the RASS reveals that 63\% of the slew sources
have an X-ray counterpart of which 68\% (90\%) lie within 16 (31) arcseconds                                                                            
(Fig.~\ref{fig:ROSAThist}).
This gives confidence that the
majority of slews have well reconstructed attitude. 
To form a sample of catalogues with highly accurate positions
but which minimise the number of false matches, we used the Astronomical Virtual Observatory (AVO)
to correlate the slew positions against non-X ray SIMBAD catalogues. This
gave 508 matches of which 68\% (90\%) were contained within 8 (17) arcseconds (Fig.~\ref{fig:Simbadhist}).

\section{The catalogue}

Source lists were produced by searching each slew down to a likelihood
$\detml>8$ and combined to produce an initial catalogue of 5180 
detections. This is available as the 'full' catalogue where known spurious
detections have been flagged out according to a set of criteria
laid out in the next section.

\subsection{Causes of spurious detections}
\label{sect:flags}
\subsubsection{Multiple detections of an extended source}
\begin{figure}
\begin{center}
\begin{tabular}{c}
\rotatebox{-90}{\includegraphics[height=5cm]{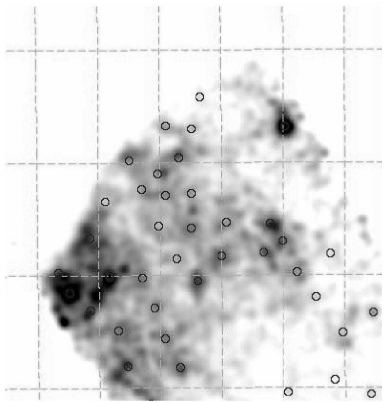}}
\end{tabular}
\end{center}
\caption[Large extended sources: Puppis-A]
{ \label{fig:pupa} The slew image of the large SNR, Puppis-A. It is 
detected as many small sources (circles).
}
\end{figure}

The source detection software attempts to parameterise source extents
up to 20 pixels (82 arcseconds here) in radius. Larger sources, or sources with
discontinuous or lumpy emission are reported as multiple small sources.
This is particularly evident for the bright supernovae remnants (SNR), 
e.g. Puppis-A
which results in 81 separate, confused, detections (Fig.~\ref{fig:pupa}).
 All affected sources
have the VER\_INEXT flag set true.

\subsubsection{The wings of the PSF of a bright source}
It was noticed during the construction of the 1XMM serendipitous 
source catalogue
that, due to the imperfect modelling of the PSF, a halo of false detections
is often seen around bright sources. The same effect is seen in slew
exposures but due to the reduced exposure time is only important for
very bright sources $\gg 10$ c/s. All affected sources
have the VER\_HALO flag set true.

\subsubsection{High background}
\begin{figure}
\begin{center}
\begin{tabular}{c}
\rotatebox{-90}{\includegraphics[height=8cm]{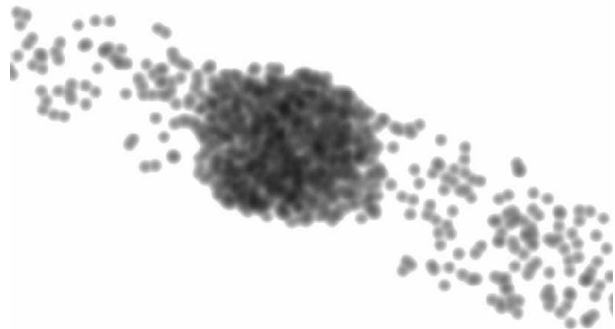}}
\end{tabular}
\end{center}
\caption[Effect of background flare]
{ \label{fig:bgndflare} A heavily smoothed section of the slew 9097100002.
The bright patch of events in the centre has been produced
by a background flare. 
}
\end{figure}

Flares in the background, due to solar protons (\cite{lumb02}, \cite{carter07}), cause a sudden
increase in the number of events seen in a slew which mimic the effect
of slewing over an SNR (Fig.~\ref{fig:bgndflare}). These flares typically last between  
10 and 40 seconds and hence affect between 15 arcminutes and 1 degree
of a slew. No automatic flare screening has been performed on the
data but light curves of all slews have been manually inspected
and 175 sources, falling within flare-affected sections have been flagged
as bad. All affected sources have the VER\_HIBGND flag set true.

\subsubsection{Bright sources outside detector field-of-view}
Reflections from the CRAB SNR, about 10 arcminutes outside the 
EPIC-pn field-of-view during the slew 9041000004, and the TYCHO SNR, 
5 arcminutes outside the field-of-view during the slew 9058600002, 
caused 5 false detections; VER\_HALO flag set true.

\subsubsection{Bad position}
Seven sources are found near the edge of the detector or the edge of an image
which leaves an uncertainty as to the true count rate of the
source and where its centre lies.
These sources have the VER\_NREDG flag set true and also the
VER\_PSUSP flag set true to indicate that the position is suspect.
In addition, sources in sections of slew with bad attitude
(see section~\ref{sect:attitude})
have their VER\_PSUSP flag set true.

\subsubsection{Zero exposure}
Two sources are flagged as false because they lie at the very edge of the 
slew and are reported as having zero exposure time in one or more 
of the energy bands. These have the VER\_FALSE flag set true.

\subsubsection{Optical Loading}
\begin{figure}
\begin{center}
\begin{tabular}{c}
\rotatebox{-90}{\includegraphics[height=8cm]{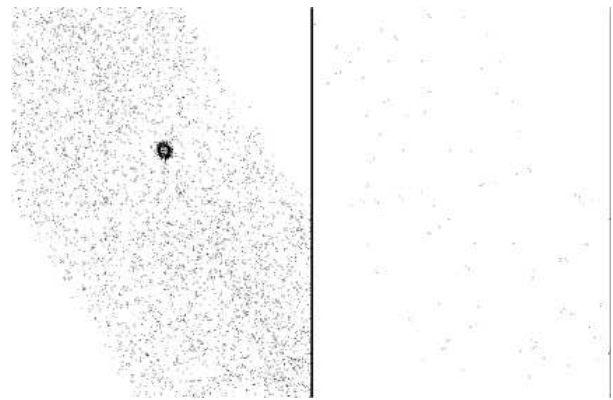}}
\end{tabular}
\end{center}
\caption[Optical loading check]
{ \label{fig:optload} Left: A raw image of gamma Cru, an M star with 
m$_{v}$=1.6, detected in slew 90130900002.  Right: The same image after
applying the filter (FLAG==0, PATTERN==0, PI$>$200).
}
\end{figure}

Despite initial concerns that the lack of a detector offset map during slews,
would lead to optical loading problems, in practice little or no effect 
was found. In Fig.~\ref{fig:optload} we show a first magnitude star
which is brilliant in raw slew data but completely disappears when the
default event filtering is applied. None of the source fluxes in the slew
catalogue are believed to be contaminated by optical photons.

\subsection{Statistical fluctuations}
\begin{figure}
\begin{center}
\begin{tabular}{c}
\rotatebox{-90}{\includegraphics[height=8cm]{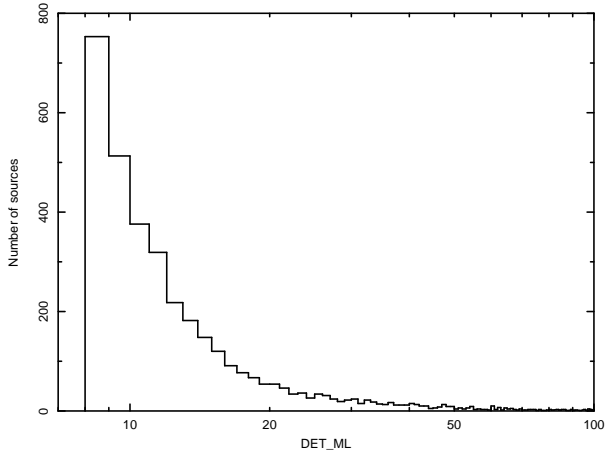}}
\end{tabular}
\end{center}
\caption[Sources as function of likelihood]
{ \label{fig:srccntvdetml} Number of detections as a function of detection likelihood
(total band; 0.2--12 keV).
}
\end{figure}

\begin{figure}
\begin{center}
\begin{tabular}{c}
\rotatebox{-90}{\includegraphics[height=8cm]{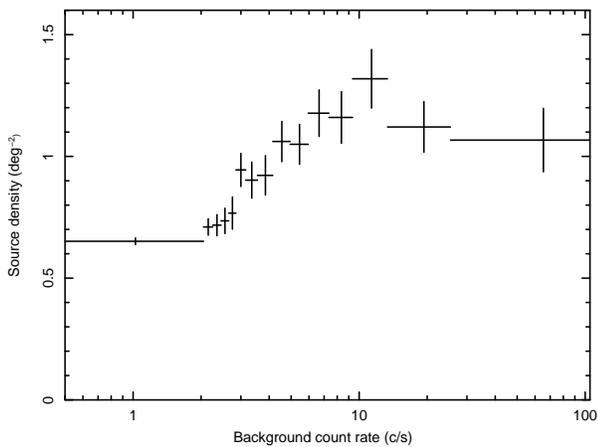}}
\end{tabular}
\end{center}
\caption[The source density as a function of the
background rate in the image in which the source was found]
{ \label{fig:srccntvbckgnd} The mean density of total band (0.2-12 keV) 
sources, flagged good in the full catalogue, plotted as a 
function of the background count rate (PI$>10000$).
}
\end{figure}

The number of detections, after removing the spurious sources 
highlighted in the previous section, rises steeply with 
decreasing detection likelihood (Fig.~\ref{fig:srccntvdetml}) as expected. 
They are also seen to have a dependency on the background rate within 
the image in which the source was found (Fig.~\ref{fig:srccntvbckgnd})
; where the background is defined as the event
count rate above 10 keV ($PI>10000$).

\begin{figure}
\begin{center}
\begin{tabular}{c}
\rotatebox{-90}{\includegraphics[height=8cm]{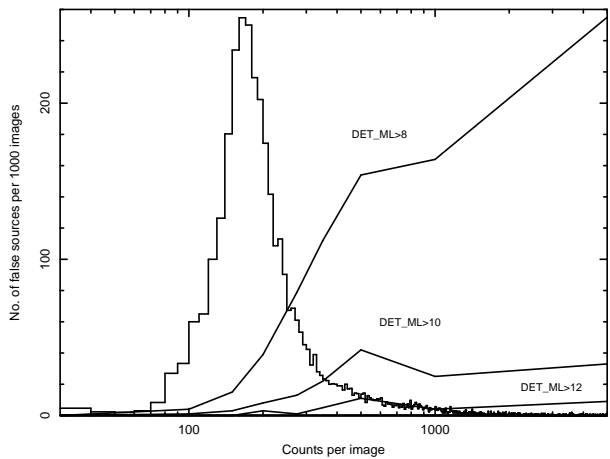}}
\end{tabular}
\end{center}
\caption[False source rate from simulations]
{ \label{fig:falserate} The false source rate as determined from simulations.
The histogram represents the distribution of counts in the real slew images.
}
\end{figure}

\begin{table}
\caption{The number of sources found in slew images and in 
simulated slew images for combinations of minimum detection likelihood
and background count rate}
\label{tab:spurFrac}      
\begin{center}
\begin{tabular}{l l l l l l}        
\hline\hline                 
ML$^{a}$ & bkg$^{b}$ & \multicolumn{4}{c}{Band$^{c}$} \\
     &  c/s & Combined & Total & Hard & Soft \\ 
\hline                        
\hline                                   
8    &    -     & 4710 / 929   & 3863 / 580 & 692 / 272 & 2606 / 186 \\
8    &  3.0  & 3451 / 456   & 3018 / 348  & 427 / 93 & 1981 / 69 \\
10    &    -     & 2998 / 195   & 2580 / 118 & 312 / 61 & 2031 / 46 \\
10    &  3.0  & 2419 / 106   & 2155 / 86 & 239 / 24 & 1638 / 13 \\
12    &    -     & 2161 / 40   & 1875 / 28 & 185 / 12 & 1661 / 10 \\
12    &  3.0  & 1782 / 21   & 1587 / 20 & 157 / 5 & 1354 / 2 \\
14    &    -     & 1700 / 9   & 1470 / 7 & 139 / 4 & 1361 / 1 \\
14    &  3.0  & 1423 / 6   & 1257 / 4 & 119 / 3 & 1122 / 1 \\
10/14$^{d}$ &  3.0/-  & 2696 / 109   & 2368 / 89 & 259 / 25 & 1877 / 13 \\
\hline
\end{tabular}
\end{center}
$^{a}$ The minimum source likelihood. \\
$^{b}$ The maximum background rate accepted within an image, defined
as the count rate of events with energy greater than 10 keV (PI$>10000$). \\
$^{c}$ The number of detected sources (first number) and the number of expected false sources (second number) from simulations, in this energy band, for this combination of
minimum detection likelihood and maximum background count rate. 
The "combined" band is made from the unique distinct sources detected in
any of the total, soft or hard bands.\\
$^{d}$ A selection of all sources with DET\_ML$>$14 and sources with 
DET\_ML$>$10 from images where the background count rate is less than 3 c/s.\\
\end{table}


Simulations have been conducted to investigate the 
relationship between the number of spurious
sources expected from background fluctuations and the number of 
events in an image. Simulated
slew images have been created by inserting events into a template
slew image, of 842 by 600 pixels and area 0.5 square degrees, at 
random positions. A flat spatial distribution of events 
has been used
because the background is likely dominated by charged particle induced
events, which show little variation across the detector, and internal
flourescent emission lines which map the distribution of metals in the
detector itself (\cite{lumb02}).
The resultant simulated images have been source searched
and reveal a strong increase in the number of spurious detections
 with image counts, that rises steeply until
 around 500 events and then flattens out (Fig.~\ref{fig:falserate}). 
We have overlaid the distribution of counts in the real slew 
images in figure~\ref{fig:falserate} 
to show that the majority of real images should generate few spurious sources
for \detml$>12$.

To assess the absolute number of false sources found by the source detection
chain as a function of \detml, the positions of the photons in all 11137
slew images have been randomised and the images source-searched again.
Results show that a significant number of sources with low \detml  can
be expected to be false. From Fig.~\ref{fig:falserate} we know that 
the false source rate is influenced by the number of events in the image 
which is in turn 
related to the background rate. Selections of \detml  and image background rate
can be made from the simulation results to choose a particular
spurious source fraction for a given purpose (Table~\ref{tab:spurFrac}). 
It is clear that the hard band, having
typically a higher background than the soft band and a lower signal to 
noise ratio than either the soft or total band, is most affected by
statistical fluctuations. A source selection [\detml$>14$ or
(\detml$>10$ and image\_bckgnd\_rate$<=$3.0)] gives an expected 25 false
detections from 259 hard band sources ($\sim 9\%$), 13 from 1877 (0.7\%)
 for the soft band and 109 from 2696 (4\%) for all sources. 
Note that 80\% of images have
a background rate $<=3.0$ c/s. A list of sources has been produced from this
 selection and is termed the 'clean' catalogue. Finally, four sources which 
showed large position errors, due to the attitude file problems
discussed in section~\ref{sect:attitude} had their VER\_PSUSP flag
set true in the 'full' catalogue and were removed from the 'clean' 
catalogue. This leaves 
a total of 2692 sources (257 in the hard band) in the 'clean' catalogue. 

\subsection{Released Catalogues}
Both the 'full' catalogue, with 5180 detections with \detml$>8$ and
the 'clean' catalogue, with 2692 sources having \detml$>14$ or
(\detml$>10$ and image\_bckgnd\_rate$<=$3.0) and the VER\_INEXT, VER\_HIBGND, 
VER\_HALO and VER\_FALSE flags set false (see section ~\ref{sect:flags}) 
are available from {\it http://xmm.esac.esa.int/external/xmm\_data\_acc/xsa}. 
The catalogues contain columns with the detection 
threshold, background
rate and spurious source flags sufficient to allow the user to select
a subsample for a particular scientific purpose. The 'clean' catalogue
is conservative for the soft band but may not be strict enough for some
applications in the hard band.

From hereon we discuss the properties of sources drawn from the 'clean'
catalogue unless otherwise stated.  

\section{Source properties}
\subsection{Naming convention}
The adopted name for sources detected in the XMM-Newton slew survey 
starts with the prefix, XMMSL1, and 
then encodes the J2000 sky position, e.g. XMMSL1 J010537.6+364858 . The name 
is assigned in two passes. When the three independent energy band source 
lists are combined to form one catalogue the source name is set using 
the position in the band where the DET\_ML likelihood is the highest. 
A second pass is then performed such that sources which have been 
observed in more than one slew are given the same name. 
Again, priority is given depending on the detection likelihood. 
Detections are deemed to be from the same source if their 
centres lie within 30 arcseconds of each other. 
Given the scarcity of slew sources (0.8 detections in the 'full' catalogue
 per square degree) 
on the sky, 30 arcseconds was found to be a reasonably robust match radius 
for point sources. It is not so good for extended sources 
and the catalogue in some cases contains multiple detections of the 
same extended source with different names.

\subsection{Source Extent}
\begin{figure}
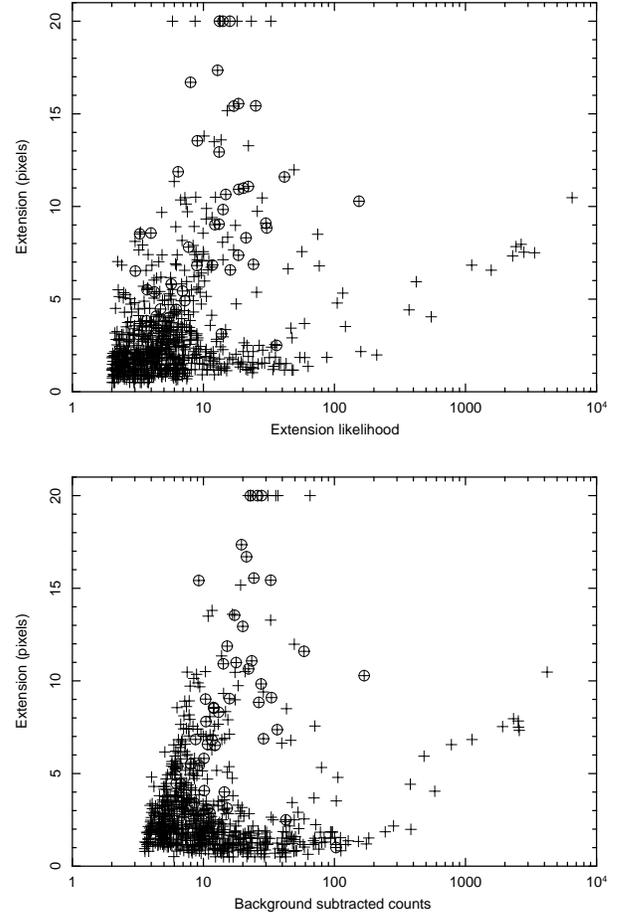

\centering
\rotatebox{-90}{\includegraphics[height=8cm]{9193fg15.ps}}
\\
\vspace{0.5cm}
\rotatebox{-90}{\includegraphics[height=8cm]{9193fg16.ps}}
\caption[Source extent v source counts]
{ \label{fig:extvcnts} Source extension, in units of 4.1 arcseconds image
pixels, plotted against the extension likelihood (upper) and number of 
source counts (lower) for extended sources detected in the total energy band.
Sources identified as clusters of galaxies are marked with circles.
}
\end{figure}

The source search algorithm attempts to parameterise detections as a 
point source or as an extended source with a radius of up to 20 pixels (82 arcseconds).
A large number of sources (30\%) are detected as being extended
(extension parameter$>1$), 
The measured extension is related to the extension likelihood and
to the number of source counts as shown in Fig.~\ref{fig:extvcnts}.
Here it can
be seen that there is an upper branch where an increasing number of 
photons are needed to detect larger source extensions and a lower 
branch where small extensions are found for a large range of source
strengths and extension likelihoods. The point spread function (PSF)
for the EPIC-pn detector is a function of off-axis angle
(the distance between the optical-axis and the source). In the data analysis 
we have used the average off-axis angle of the path of a source
through the detector, to calculate the appropriate PSF.
This is reasonably accurate for the LW and FF modes where the frame time
is short and the extension of the PSF along the slew direction small,
but introduces an inaccuracy for the $\sim20\%$ of observations taken in
eFF mode, where the extension along the slew direction is $\sim18\arcsec$
 (Table~\ref{tab:obsMode}). The lower branch is contaminated by false detections
of extension caused by this effect which can be demonstrated by the
fraction of eFF sources in this branch (35\%) compared with the expected
20\% in the upper branch. At large count rates, photon pile-up depresses the 
counts in the central pixels of the source profile also causing an apparent 
extension. A correct treatment of the slew PSF is needed to properly 
parameterise source extension, nevertheless sources falling in the upper
branch of Fig.~\ref{fig:extvcnts} are considered to be genuinely extended. 

\begin{table}
\caption{Observing mode statistics}
\label{tab:obsMode}      
\begin{center}
\begin{tabular}{l c c c}        
\hline\hline                 
Mode & Frame Time & Extension$^{a}$ & Fraction$^{b}$ \\    
          & (ms) & (arcseconds)  &  \%                \\    
\hline                        
\hline                                   
FF & 73 & 6.6 & 69 \\
eFF & 199 & 17.9 & 21 \\
LW  & 48   & 4.3   & 10 \\
\hline
\end{tabular}
\end{center}
$^{a}$ The extension of the PSF along the slew direction caused by the
satellite movement during CCD integration. \\
$^{b}$ The percentage of the slew sky area covered in this mode. \\
\end{table}

\subsection{Count rates}
\label{sect:countrates}
The count rates are calculated from the background subtracted counts within 
a circle about the source, corrected for the
encircled energy fraction and divided by the PSF-weighted exposure time
within the source region (see the {\em emldetect} user guide for more details
 \footnote{http://xmm.esac.esa.int/sas/6.1.0/doc/emldetect/index.html}).
For reasons of limited resources, source searching was performed in
all bands using the exposure map for the total energy band. This produces 
incorrect exposures in the soft and hard bands due to the energy-dependent
vignetting. A correcting factor:

\begin{equation}
e_{\mathrm{h}} = 0.0013806 \times e_{\mathrm{t}}^{3} + 0.0085632 \times e_{\mathrm{t}}^{2} + 0.84282 \times e_{\mathrm{t}}
\end{equation}
\begin{equation}
e_{\mathrm{s}} = 0.0014723 \times e_{\mathrm{t}}^{3} - 0.058884 \times e_{\mathrm{t}}^{2} + 1.3509\times e_{\mathrm{t}}
\end{equation}

where e$_{s}$, e$_{h}$ and e$_{t}$ are the soft, hard and total band
exposure times respectively, has been applied to the exposure times
to correct for this effect. These factors were calculated by comparing
several sets of total, soft and hard-band exposure maps and fitting a function 
to the relationship between the bands. 
This introduces a systematic uncertainty
of $\sim5$\% into the soft and hard band exposure times and count rates.
Due to the limit of 82 arcseconds on source extent used by the search algorithm, sources extended beyond this value will have their count rate underestimated.
Very bright sources are affected by photon pile-up which tends to reduce
the count rate. The source strength limits for the observing modes are 
given by \cite{struder} for FF mode as 6 c/s and for LW mode as 9 c/s.
For eFF mode the pile-up limit is in principle 2 c/s for a pointed observation
but will be higher here as the slewing movement, together with the relatively 
long frame time, will reduce the count rate
on the central pixel of the PSF.

\begin{figure}
\begin{center}
\begin{tabular}{c}
\rotatebox{-90}{\includegraphics[height=8cm]{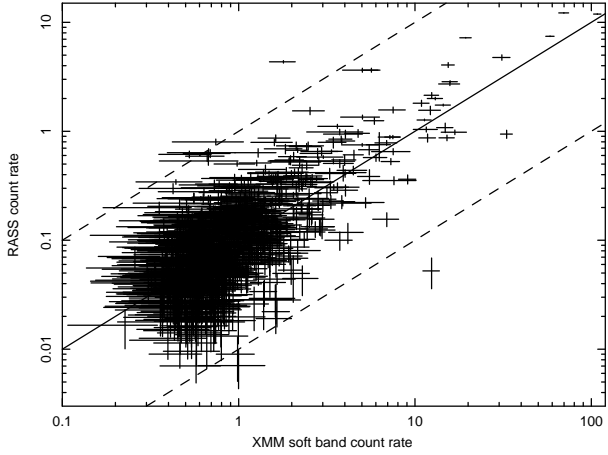}}
\end{tabular}
\end{center}
\caption[A comparison of RASS and slew soft band count rates]
{ \label{fig:roscnts} A comparison of RASS and slew soft band count rates 
(c/s). The
solid line represents a ratio of 10:1 and the dotted lines represent
a factor of 10 variation from this value.
}
\end{figure}

A comparison of the soft band count rates against RASS count rates
is presented in Fig.~\ref{fig:roscnts} for 894 non-extended sources 
with a ROSAT counterpart. The count rate ratio,
XMM/RASS, is typically $\sim10$ but varies considerably with the
source spectrum. The comparison of these two surveys, coupled with
upper-limits analysis represents a powerful tool for finding high 
variability X-ray sources. An initial analysis of high variability extragalactic
sources found in this way has been published in \cite{esquej}.

\subsection{Flux conversion}

Source fluxes have been calculated from count rates based on energy conversion factors assuming a spectral model of an absorbed power-law with $N_{H} = 3 \times 10^{20}$ cm$^{-2}$ and photon index 1.7 (see XMM science survey centre memo,
SSC-LUX TN-0059, for a general description of the technique). The energy 
conversion factors are given in Table~\ref{tab:fluxConv}.
 
\begin{table}
\caption{Flux conversion factors}
\label{tab:fluxConv}      
\begin{center}
\begin{tabular}{c c c}        
\hline\hline                 
Band & energy range & Conversion factor$^{a}$ \\    
          & (keV) &                    \\    
\hline                        
\hline                                   
Total & 0.2--12.0 & 3.16 \\
Hard  & 2.0--12.0 & 9.14 \\
Soft & 0.2--2.0 & 1.44 \\
\hline
\end{tabular}
\end{center}
$^{a}$ Converts from source count rate (c/s) to flux in the given energy 
band in units of $10^{-12}$ ergs/s/cm$^{2}$ \\
\end{table}

The soft-band fluxes are particularly dependent on the spectral model
used and can be quite discrepant for stars where the absorbing
column may be small. 

At the fluxes probed here, source confusion within the 20 arcseconds  extraction 
radius is almost absent.

\subsection{Hardness ratio}
In Fig.~\ref{fig:allsrchr} we plot the positions in Galactic coordinates of the soft
and hard band sources and separately all the sources, colour-coded by 
hardness ratio. The hardness ratio is defined as 

    \begin{equation}
      H_{R} = ( S_{H} - S_{S} ) / ( S_{H} + S_{S} )
    \end{equation}

where $S_{H}$ is defined as the source count rate in the hard band
and $S_{S}$ as the count rate in the soft band.

As expected, hard sources, notably the bright LMXB, predominate in the 
Galactic plane and soft sources at higher Galactic latitudes.

\begin{figure}
\begin{center}

\rotatebox{90}{\includegraphics[bb=25 -675 500 -50, height=9cm]{9193fg18.ps}}

\rotatebox{90}{\includegraphics[bb=25 -675 500 -50, height=9cm]{9193fg19.ps}}

\rotatebox{90}{\includegraphics[bb=25 -675 500 -50, height=9cm]{9193fg20.ps}}
\end{center}
\caption[An AITOFF projection in Galactic coordinates of the slew
sources.]
{ \label{fig:allsrchr} An AITOFF projection in Galactic coordinates of 
sources from the first \xmm slew survey, where the circle size scales 
logarithmically with the count rate. Top panel: the total band sources;  
the hardness ratio is colour-coded such 
that light red is soft and blue is hard. Middle panel: the soft band
sources. Bottom panel: the hard band sources. The two bright sources seen
in the hard-band plot at 
latitude $\sim -30$ are LMC~X-1 and LMC~X-3.
}
\end{figure}

\subsection{Spectra}
Spectral analysis of slew data is fundamentally complicated by 
the fact that a source bright enough to produce a reasonable spectrum 
will suffer from photon pile-up. 
A quantative analysis of these spectra awaits the implementation of a
realistic pile-up correction and the development of an energy-dependent 
point spread function, which will be a vignetting-weighted average of
the PSF at each detector position traversed by the motion of the source
through the detector.  

\section{Sample properties}

\begin{figure}
\begin{center}
\begin{tabular}{c}
\rotatebox{-90}{\includegraphics[height=9cm]{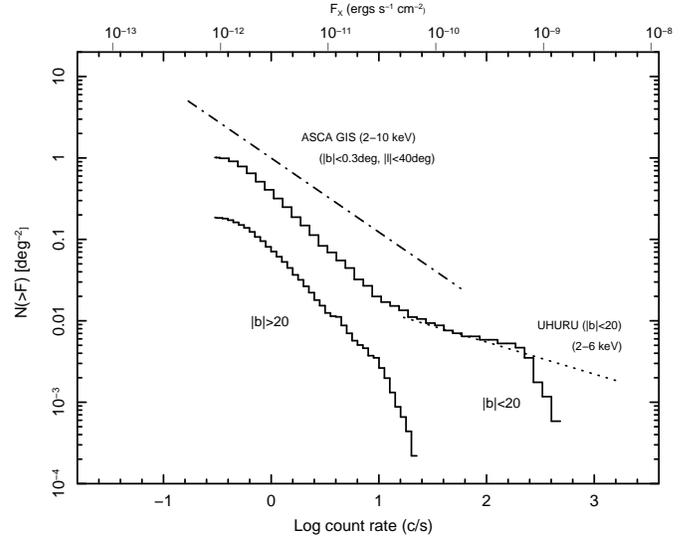}}
\end{tabular}
\end{center}
\caption[LogN/logS plot for total band sources]
{ \label{fig:logns} The cumulative number count distribution of total band (0.2-12 keV) sources inside and outside the Galactic plane, plotted against count rate and absorbed flux. A spectral model of an absorbed power-law of photon index 1.7
and $N_{H} = 3 \times 10^{20}$ cm$^{-2}$ has been used to convert the slew count rate into flux.
The UHURU Galactic plane (dotted line; \cite{forman})
and the ASCA Galactic plane survey (dotdash line; \cite{sugizaki}) log {\it N}--log {\it S} relations
have been displayed for comparison. In both of these cases the fluxes have been converted to an energy range 0.2--12 keV and to the above spectral model.
}
\end{figure}

In Fig.~\ref{fig:logns} we show the cumulative number count distribution for the
full energy band for sources within ($|b|\le20^\circ$) and outside 
($|b|>20^\circ$)
the Galactic plane. A linear fit to sources outside the Galactic plane,
in the central part of the distribution (log counts between -0.2 and 1.0) 
gives a slope of $-1.41\pm0.01$. This is a little flatter than the
Euclidian slope of 1.5 found by ASCA (\cite{ueda}) and earlier by
HEAO-1 A2 (\cite{piccinotti}) and Ginga (\cite{kondo}).
The difference in slope may be due to incompletenes, which appears to become evident in 
the source distribution below $\sim 0.8$ c/s. 
The distribution of sources within the 
Galactic 
plane shows a break at about 12 c/s (F$_{0.2-12}=3.6\times10^{-11}$ 
ergs s$^{-1}$ cm$^{-2}$ for an absorbed power-law with $N_{H} = 3 \times 10^{20}$ cm$^{-2}$ and photon index 1.7). At fainter fluxes the slope
is $-1.29\pm0.02$ while for brighter sources it is $-0.43\pm0.02$.
The bright source population agrees well with the previous result of UHURU 
(\cite{forman}), except at fluxes above F$_{0.2-12}\sim 8\times10^{-10}$ ergs s$^{-1}$ cm$^{-2}$ where gross pile-up effects significantly reduce the observed count rate (see section~\ref{sect:countrates}). 
The log {\it N}--log {\it S} relation below F$_{0.2-12}=3.6\times10^{-11}$
is steeper than the $0.79\pm0.07$ observed by ASCA in a survey with 
$|b|<0.3^\circ$ in the 2--10 keV band (\cite{sugizaki}). It is likely that 
the difference is at least partly due to the sky regions sampled. 
The wider \xmm slew sample will
, for example, contain a higher fraction of extragalactic sources. 
A thorough analysis of the log {\it N}--log {\it S} relation will require a
careful analysis of the effects of pile-up, completeness, sky coverage
 and the Eddington bias and we defer this to a later date when a larger
fraction of the sky has been covered. 


\section{Duplicate detections and variability}

In the full catalogue 96 sources have been detected in more than one slew.
Count rate variability up to a factor 5, 5 and 2 is seen in the total, soft 
and hard bands respectively. The greatest variability is seen in the total
and soft band
fluxes of XMMSL1 J125616.0-114632 (2MASXJ12561595-1146367); which 
varied by a factor 5 between slews 9037400004 and 9083600004 (a baseline
of 2.5 years). An upper
limits analysis will likely find greater variability.

\section{Identifications}
All sources detected in the survey have been correlated with different 
catalogues in order to identify the XMM-Newton slew sources with 
previously known objects (See Table~\ref{tab:cats} for a summary of 
the resources used). The catalogues used for this aim comprise two 
astronomical databases, a catalogue of clusters of galaxies and nine other 
catalogues (some which have been queried through the HEASARC astronomical 
database). Although the astrometric uncertainty of the slew sources was found 
to be 8 arcseconds, the offset radius for the correlations was 
30 arcseconds (with a few exceptions described below) in order 
to include sources from catalogues with worse accuracy or truncated coordinates. 
For the EXOSAT CMA catalogue the offset radius was 45 arcseconds, 
and for the Einstein IPC it was 2 arcmin, both due 
to the larger uncertainty in source coordinates. A radius of 5 
arcmin was chosen for the clusters catalogue due to the extension of 
this type of object. 
The identification process results in unidentified sources, sources with a 
single counterpart and sources with multiple matches. A hierarchical 
selection scheme has been applied for sources with different counterparts. A 
decision has been made to derive the most plausible identification candidate 
using the technique described below. Firstly, the SIMBAD and 
NED astronomical databases have been used for the cross-correlation and 
results from both databases have been compared in detail (SIMBAD has 
been queried in the frame of the Astronomical Virtual Observatory (AVO)).
When a source has the same counterpart in both databases the one 
selected for the identification is that which gives more specification about 
the source category. When contradictory identification candidates have 
been found, the one with the smallest positional offset has been chosen. 
These two 
databases provide the large majority (90\%) of the total number 
of identifications. Then, a correlation with a clusters table 
(Abell and Zwicky Clusters of Galaxies ) has been performed. 
The final identification for sources with a reported extension greater
than 2 pixels, which have a SIMBAD/NED and also a cluster counterpart,
comes from the 
clusters table. The remaining catalogues used for the cross-match are 
listed below ordered in priority for the preferred identification. For sources 
with multiple matches in a catalogue, the identification was selected as 
the closest match. These catalogues are: All-Sky Optical 
Catalog of Radio/X-Ray Sources, Catalog of PSPC 
WGA Sources, Einstein IPC Sources Catalog, EXOSAT CMA Images/Lightcurves, 
ROSAT All-Sky Survey catalogue, ROSAT Results Archive Sources for 
the PSPC, ROSAT Results Archive Sources for the HRI, RXTE Master 
Catalog, XMM-Newton Serendipitous Source Catalog, Version 1.1.0, 
INTEGRAL Bright Source Catalog. Results from the identification process 
appear in the final catalogue in the columns: 

\begin{table}
\caption{Catalogues used for source identification}
\label{tab:cats}      
\begin{center}
\begin{tabular}{l c l}        
\hline\hline                 
Catalogue & match radius & Reference \\    
          & (arcseconds) &                    \\    
\hline                        
\hline                                   
SIMBAD    & 30       & CDS$^{a}$ \\
NED       & 30       & NED$^{b}$ \\
Abell clusters & 300 & \cite{Abell} \\
Zwicky clusters & 300   & \cite{zwicky}\\
All-Sky Optical Catalog & & \\
of Radio/X-Ray Sources & 30 & \cite{fleschAndHard}\\
Catalog of PSPC WGA sources & 30 & \cite{white04}\\
Einstein IPC & 120 & \cite{harris}\\
EXOSAT CMA & 45 & HEASARC$^{c}$\\
RASS & 30 & \cite{voges} \\
ROSAT PSPC Results Archive & 30 & HEASARC$^{c}$\\
ROSAT HRI Results Archive  & 30 & HEASARC$^{c}$\\
RXTE Master Catalog & 30 & HEASARC$^{c}$\\
1XMM V1.1.0 & 30 & \cite{watson}\\
INTEGRAL BSC & 30 & HEASARC$^{c}$\\
\hline
\end{tabular}
\end{center}
$^{a}$ CDS, Strasbourg (2006) \\
$^{b}$ The NASA/IPAC Extragalactic Database (NED) is operated by the Jet Propulsion Laboratory, California Institute of Technology, under contract with the National Aeronautics and Space Administration.\\
$^{c}$ Data obtained from the High
Energy Astrophysics Science Archive Research Center (HEASARC), provided
by NASA's Goddard Space Flight Center.

\end{table}

IDENT: name of the source \\

ALTIDENT: alternative name of the source \\

ID\_DIST: distance in arcminutes between the slew source and the identification.  Distances have been rounded to the nearest 0.1 arcminutes to ensure a uniform accuracy across the catalogues.  \\

ID\_CATEGORY: type of the identified source that, when existing, 
has been extracted from the catalogues (it is not very homogeneous 
because type convention differs between catalogues) \\

RASSNAME: the closest RASS match \\

RASSDIST: distance in arcseconds to the closest RASS match.

Of the full catalogue sources, not flagged as spurious, 51\% are identified.
The fraction rises to 71\% when only sources from the CLEAN catalogue are
considered. Of these, 48\% are extragalactic, 30\% are galactic and the
remainder are of unknown type, e.g. "X-ray source".

The identification list is expected to improve as more catalogues come on-line
and with follow-up observations. The list is maintained at 
{\it http://xmm.esac.esa.int/external/xmm\_science/slew\_survey}
{\it/ident\_tab.html}
and suggestions for counterparts are welcomed.

\begin{figure}
\begin{center}
\begin{tabular}{c}
\rotatebox{-90}{\includegraphics[height=9cm]{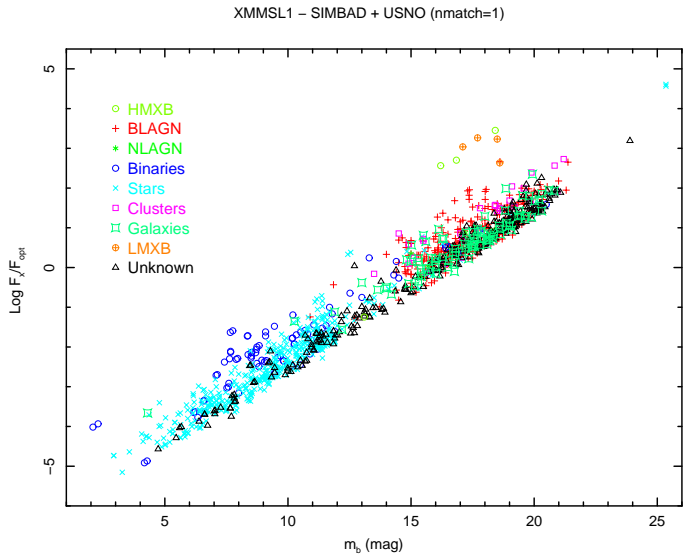}}
\end{tabular}
\end{center}
\caption[The fx/fopt ratio against optical magnitude]
{ \label{fig:magflux} The log of the ratio of the X-ray to optical flux
plotted against the optical magnitude of the counterpart.
}
\end{figure}

In figure~\ref{fig:magflux} we plot the ratio of the X-ray to optical flux against the optical magnitude. Here we define the X-ray to optical flux ratio
using the total band X-ray flux and the optical blue-band flux. Optical 
fluxes have been obtained either from the Simbad counterpart or
from the USNO B1 magnitude for the cases where a single plausible match 
is found within the slew error circle. Distinct source types lie in 
distinct regions of the
plane and statistically it can be seen that the unknown
sources with an unambiguous optical counterpart predominantly fall
in the region occupied by AGN but there are also a significant fraction of 
sources which lie in the region occupied by stars and binaries that 
have not been assigned a source category.

\begin{figure}
\begin{center}
\begin{tabular}{c}
\rotatebox{-90}{\includegraphics[height=9cm]{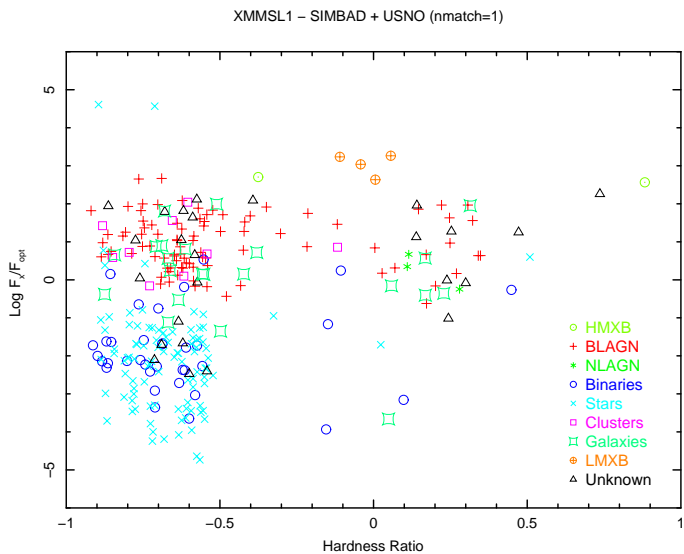}}
\end{tabular}
\end{center}
\caption[The ratio of X to optical fluxes plotted against the hardness ratio]
{ \label{fig:hrmag} The ratio of X to optical fluxes plotted against the X-ray hardness ratio. In the case of a non-detection in the hard or soft band, the hardness
ratio has been calculated by assuming an upper limit of three photons in
the non-detection band. For reasons of simplicity neither error-bars nor
upper limit arrows are shown on the plot.
}
\end{figure}

Source types may be further distinguished by considering the X-ray
hardness ratio (HR; e.g. \cite{delaceca}). In general the low number of counts 
precludes an accurate measurement of the HR for slew sources. 
In figure~\ref{fig:hrmag} we show sources which have either a detection 
in the hard band or $\ge10$ counts in the soft band and use an upper limit
of three photons in the case of a non-detection in one of the bands. 
The two points with the
highest X-ray to optical ratio are two detections of the isolated neutron star
RXJ~1856.6-3754. Unidentified sources are present in both the hard and soft
regions of the plot.

\section{Summary}
The \xmm slew data represent the deepest near all-sky hard band X-ray survey made to date, while the soft band survey is comparable with the RASS.
The source density, from the clean catalogue, is $\sim0.45$ per square
degree and $\sim70\%$ have plausible identifications. 
The \xmm slew survey catalogue will continue to grow as the mission
continues and it is expected that a sky coverage in excess of 50\% 
will eventually be achieved.
With the excellent attitude reconstruction 
this will leave a powerful legacy for future variability studies in
the soft and hard X-ray bands. 

\acknowledgements

We would like to thank Mark Tuttlebee, Pedro Rodriguez and Aitor Ibarra 
for their patient explanations of how \xmm performs slews.
We thank Ramon Munoz and his team for the compilation of the slew datasets
and Mike Watson and Norbert Schartel for many useful discussions.
This research has made use of the SIMBAD database,
operated at CDS, Strasbourg, France the NASA/IPAC Extragalactic 
Database (NED) which is operated by the Jet Propulsion Laboratory, 
California Institute of Technology, under contract with the National 
Aeronautics and Space Administration and data obtained from the High 
Energy Astrophysics Science Archive Research Center (HEASARC), provided 
by NASA's Goddard Space Flight Center.
The XMM-Newton project is an ESA science mission with instruments and contributions directly funded by ESA member states and the USA (NASA).
The \xmm project is supported by the Bundesministerium f\"{u}r Wirtschaft und Technologie/Deutches Zentrum f\"{u}r Luft- und Raumfahrt (BMWI/DLR, FKZ 50 OX 0001),
the Max-Planck Society and the Heidenhain-Stiftung. AMR acknowledges the
support of PPARC funding and PE the support of MPE funding.

\end{document}